\documentclass[conference]{IEEEtran}

\usepackage{epsfig}
\usepackage{amsmath,amssymb,amsfonts}
\usepackage{adjustbox}
\usepackage{cite}
\usepackage{todonotes}
\usepackage{hyperref}
\usepackage{graphics}
\usepackage{xcolor}
\usepackage{fancyhdr}
\newcommand{\comment}[1]{}

\hypersetup{
    colorlinks=true,
    linkcolor=blue,
    filecolor=magenta,      
    urlcolor=cyan,
    pdftitle={ConSep}
    }

\let\OLDthebibliography\thebibliography
\renewcommand\thebibliography[1]{
  \OLDthebibliography{#1}
  \setlength{\parskip}{0pt}
  \setlength{\itemsep}{0pt plus 0.3ex}
}

\fancypagestyle{FirstPage}{
\lfoot{979-8-3503-3959-8/23/\$31.00 \copyright2023 IEEE} 
}

\begin{document}\sloppy

% Title.
% ------
\title{ConSep: a Noise- and Reverberation-Robust Speech Separation Framework by Magnitude Conditioning}
%
% Single address.
% ---------------
\author{\IEEEauthorblockN{ Kuan-Hsun Ho }
\IEEEauthorblockA{\textit{Department of Computer Science and} \\ \textit{Information Engineering} \\
\textit{National Taiwan Normal University }\\
Taipei, Taiwan \\
Email: jasonho610@ntnu.edu.tw}
\and
\IEEEauthorblockN{ Jeih-weih Hung }
\IEEEauthorblockA{\textit{Department of Electrical Engineering} \\
\textit{National Chi Nan University }\\
Nantou, Taiwan \\
Email: jwhung@ncnu.edu.tw}
\and
\IEEEauthorblockN{ Berlin Chen }
\IEEEauthorblockA{\textit{Department of Computer Science and} \\ \textit{Information Engineering} \\
\textit{National Taiwan Normal University}\\
Taipei, Taiwan \\
Email: berlin@ntnu.edu.tw}
}

\maketitle

\begin{abstract}
Speech separation has recently made significant progress thanks to the fine-grained vision used in time-domain methods. However, several studies have shown that adopting Short-Time Fourier Transform (STFT) for feature extraction could be beneficial when encountering harsher conditions, such as noise or reverberation. Therefore, we propose a magnitude-conditioned time-domain framework, ConSep, to inherit the beneficial characteristics. The experiment shows that ConSep promotes performance in anechoic, noisy, and reverberant settings compared to two celebrated methods, SepFormer and Bi-Sep. Furthermore, we visualize the components of ConSep to strengthen the advantages and cohere with the actualities we have found in preliminary studies.
\end{abstract}
\begin{IEEEkeywords}
speech separation, reverberation, cross-domain, multi-resolution, magnitude, conditioning
\end{IEEEkeywords}

\section{Introduction}
\thispagestyle{FirstPage}

SPEECH separation is a specific scenario of the source separation problem, where the targets are the overlapping speech signal sources while other irrelevant signals are treated as interferences.
\comment{The interferences might contain background noise, music, conversations, or reverberation.}
Recently, the field of speech separation has been revolutionized with the advent of deep learning techniques. The previous works toward an anechoic environment \cite{dc, pit, convtasnet, dprnn, wavesplit, dptnet, sepformer1} have been fruitful and inspiring. Current systems rely, in large part, on the prestigious EMD structure, which is composed of the Encoder, Mask estimator, and Decoder. However, the assumption of an anechoic environment might be unrealistic in most speech separation studies. In practice, speech usually coincides with various interferences, and speech separation under reverberant situations is incredibly challenging \cite{whamr}.

Interestingly, we have observed opposing performances in \cite{a2r} and \cite{sepformer2}. In \cite{a2r}, the experimental results show that a smaller-sized SepFormer (SepFormer-s), i.e., the SepFormer with fewer Inter- and Intra-Transformers, could not perform well under reverberant conditions despite still giving a competitive result under an anechoic condition. After optimizations toward reverberant conditions, the resulting SepFormer is marginally superior to the optimized PIT-BLSTM. Moreover, it fails to be competitive in an anechoic condition. (See [9, Tab. 6]) Meanwhile, \cite{sepformer2} expands the work in \cite{sepformer1} on a regular-sized SepFormer and claims good results on multiple conditions. One explanation may be the advantage of exploiting more parameters, which enables the model to represent more complicated functions than the one with fewer parameters. Nonetheless, this evidences that efforts to make SepFormer a more distilled yet versatile model need further investigation.

One endeavor that adapts SepFormer to multiple conditions is Bi-Sep \cite{bisep}. Bi-Sep leverages two parallel encoders with different time resolutions and a Bi-projection Fusion (BPF) \cite{bpf} module to integrate information from different domains. However, Bi-Sep has a potential shortcoming. Despite exhibiting better performance than existing models when facing complicated environments, Bi-Sep inevitably inherits the degradation when substituting the learnable encoder with the Short-Time Fourier Transform (STFT). Moreover, although the BPF module helps determine whether the mask estimator should attend more on a shorter or longer frame, it could cause the mask estimator to have difficulty learning from two domains when encountering a simpler environment. 

Therefore, we propose a novel speech separation framework, ConSep, which exploits conditioning on magnitude spectrogram to avoid any domain mismatch or confusion. A better conditioning method could facilitate the knowledge injection instead of concatenating all the features as Bi-Sep does. ConSep likewise embraces two encoders with different time resolutions to retrieve complementary characteristics. However, we merge the respective output features into the same dominant domain, in this case, the time domain. More precisely, we modulate the time signals by magnitude spectrogram as this modulation enables the mask estimator to better distinguish speech parts and interferences during separation. Experiments show that ConSep surpasses SepFormer under an anechoic condition and prominently upgrades SepFormer under more complicated situations. This result matches our goal of enabling a model to possess the generalizability on multiple conditions.

\comment{The remainder of the paper is composed as follows. Section 2 provides our findings in the anechoic and reverberant speech separation fields. In Section 3, we explain each component of the presented method. Afterward, Section 4 presents the experimental settings of the datasets and the model, while Section 5 presents the results, analyses, and visualization. Finally, the conclusion is drawn in Section 6.}

\section{Findings}

\begin{table}[tb]
\caption{Best model configuration from references that adapt EMD models to multiple conditions and the original SepFormer.}
\begin{center}
\label{table:findings}
\begin{adjustbox}{width=8.5cm, center}
\begin{tabular}{c|ccc} \hline
\textbf{Condition(s)}          & \multicolumn{3}{c}{\textbf{Attributes}}                    \\ \hline
\textit{Anechoic}    &\textit{Loss function} &\textit{Granularity (ms)} & \textit{Encoder/Decoder pair} \\ \hline
SepFormer\cite{sepformer1}  & SI-SDR              & 2                & Learnable          \\
Conv-TasNet\cite{demy}      & SI-SDR              & 0.5              & Learnable$^3$      \\ \hline
\textit{Reverberant} &\textit{Loss function} &\textit{Granularity (ms)} & \textit{Encoder/Decoder pair} \\ \hline
SepFormer-s\cite{a2r}   & th-SDR              & 64               & STFT$^4$           \\
SepFormer\cite{sepformer2}  & SI-SDR              & 2                & Learnable          \\
Conv-TasNet\cite{demy}      & SI-SDR              & 8                & STFT$^3$           \\ 
DenseUNet-TCN\cite{compensate}& L1$^1$            & 25               & STFT               \\
WD-TCN \cite{wdtcn}    & SI-SDR            & 2               & Learnable               \\ 
Bi-Sep32 \cite{bisep}    & SI-SDR            & 2/32$^2$               & Both               \\ \hline
\multicolumn{4}{l}{
\footnotesize$^1$ L1 Loss on real, imaginary and magnitude.} \\
\multicolumn{4}{l}{
\footnotesize$^2$ Frame length of 2 and 32 ms for Learnable and STFT, respectively.} \\
\multicolumn{4}{l}{
\footnotesize$^3$ Performance reduces by less than 2 (dB) if altered to the other pair.} \\
\multicolumn{4}{l}{
\footnotesize$^4$ Performance reduces by less than 1 (dB) if altered to the other pair.} \\
\end{tabular}
\end{adjustbox}
\end{center}
\end{table}

To ascertain what fosters a model to cope with various environments, we list the optimal model configurations of relevant studies that adapt EMD models to more complicated conditions in Tab.~\ref{table:findings}. We notice that an almost unanimous consensus is to employ the time-domain loss, including SI-SDR \cite{sisdr} and th-SDR \cite{thsdr}. The time-domain loss has been proven beneficial in numerous works, even when using STFT as an encoder \cite{a2r, demy}. However, the granularity designated by each work varies over a wide range, from 0.5 ms to 64 ms. The time-domain methods usually work on short frames, whereas traditional STFT frame sizes are set to be larger (around 32 ms). As for the type of encoder-decoder pair, \cite{demy} and \cite{gtf} have argued that it is not the crucial factor to success. 

Furthermore, through our preliminary studies, we discover four actualities:
\begin{itemize}
\setlength\itemsep{0 mm}
\item[1)]Time-domain methods usually perform better in SI-SDR and worse in PESQ than STFT methods \cite{compensate, explore, sesep}. 
\item[2)] A large enough window size is mandatory to avoid contravening the prerequisite of Multiplicative Transfer Function Approximation (MTFA) \cite{a2r, mtfa}. On the contrary, recent works have gained success due to the fine-grained window size \cite{demy}. 
\item[3)] Employing STFT representation exhibits optimal performance in reverberation. However, employing the learnable encoder/decoder prevails under an anechoic condition \cite{a2r, demy}. 
\item[4)] The phase becomes uninformative within a relatively large window size \cite{phase0, phase1, phase2}.
\end{itemize}

Although some actualities seem contradictory, a reasonable scheme to incorporate all the beneficial characteristics can compensate for those contradictions. This motivates us to propose ConSep. To encourage more instantiations in the future, we plan to provide publicly available codes used in our experiments. 

\section{Our ConSep}

The high-level description of ConSep is identical to the EMD structure. Initially, the encoder transforms the mixture $x\in \mathbb{R}^T$, which contains audio from $K$ active speakers and interferences, into a representation $w$ that characterizes the signal. Then the mask estimator produces $K$ masks $\{m_k\}$ for each active speaker in the mixture. Finally, the decoder reconstructs the separated $K$ source signals in the time domain, each represented by $\hat{s}_k\in \mathbb{R}^T$.

\subsection{Encoder}

The encoder of ConSep is composed of four modules: Learnable encoder, STFT, Multi-Channel Attention (MulCA), and Modulator. 

\comment{
\begin{figure}[tb]
\centering{\includegraphics[width=8.5cm]{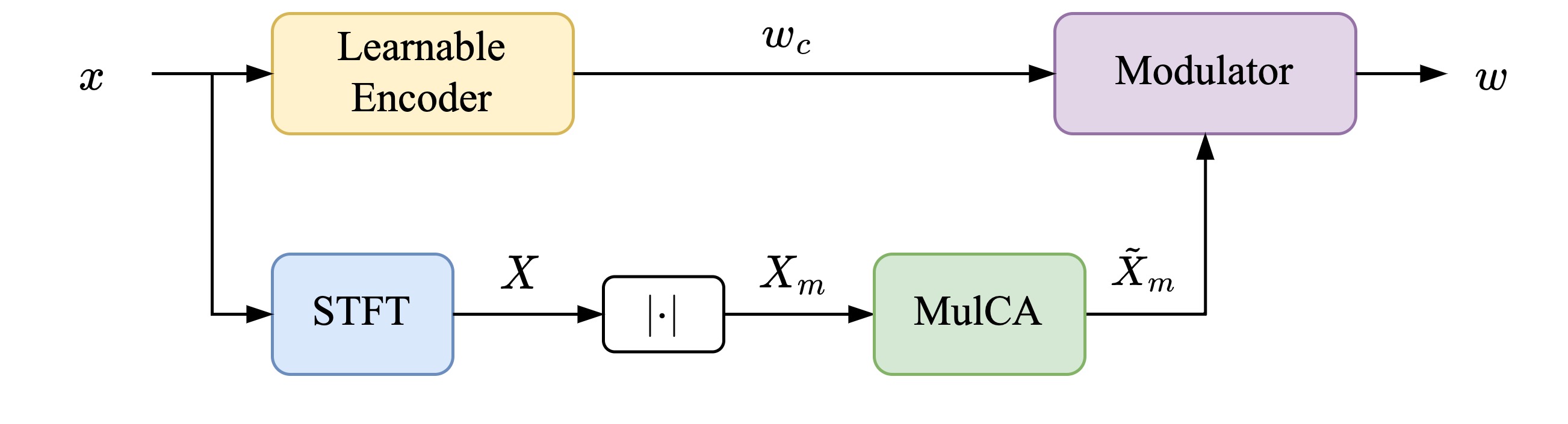}}
\caption{The encoder of ConSep.}
\label{fig:enc}
\end{figure}
}

\comment{
\begin{figure}[tb]
\centering{\includegraphics[width=8.5cm]{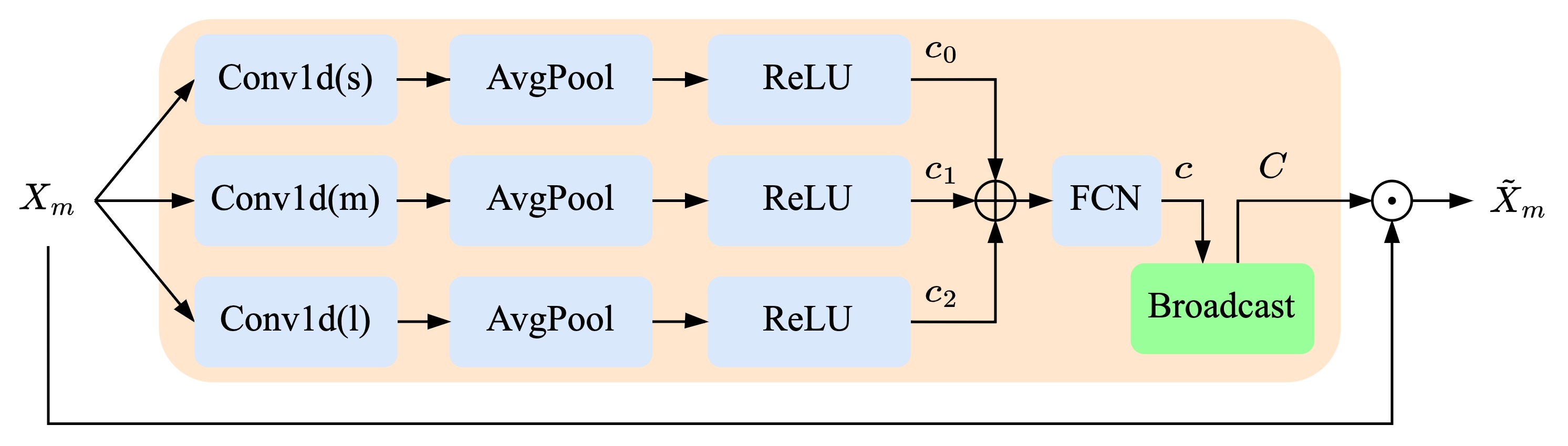}}
\caption{MulCA.}
\label{fig:mulca}
\end{figure}
}

\subsubsection{Learnable encoder}
The typical method uses one-dimensional convolutional layers (Conv1d) followed by a rectified linear unit (ReLU). %enforcing a non-negativity constraint. 
This encoder extracts the time representation $w_c\in \mathbb{R}^{N \times L}$ from the mixture $x$:

{\renewcommand\baselinestretch{0.8}\selectfont
\begin{equation}
w_c = \text{ReLU}(\text{Conv1d}(x)).
\end{equation}
\par}

\subsubsection{STFT with MulCA}
The conventional method obtains the time-frequency representation $X\in \mathbb{C}^{F \times L}$, a.k.a spectrogram, through STFT: %, which is essentially the discrete Fourier transform (DFT) to windowed and overlapped frames:
 
\begin{equation}
X[f,l]=\sum_{n=0}^{W-1} x[n+Hl]w[n]e^{-j\frac{2\pi fn}{W}},
\end{equation}
where $f$, $l$, and $n$ are the frequency bin, frame, and local time indices; $W$ is the window length, and $H$ is the hop size. After, we extract the magnitude part $X_m \in \mathbb{R}^{F\times L}$ from the spectrogram $X$. Prior to the modulation, we add a MulCA block \cite{adafsn, fsn+} to weigh the magnitude spectrogram $X_m$. MulCA regards different frequency bins as channels, giving them different weights using Channel Attention. 

The intuition behind MulCA is that the energy of a speech utterance usually distributes non-uniformly in frequencies, and different frequency components are unequally crucial to human perception \cite{mulca1, mulca2}. For example, the lower frequency band tends to contain high energies, tonalities, and long-duration sounds; the higher frequency band may have low energies, noise, and rapidly decaying sounds. The following equations express the operations of a MulCA:
\begin{equation}
    \begin{gathered}
    c_i = \text{ReLU}(\text{AvgPool}(\text{Conv1d}(X_m; k_i))), i=0, 1, 2, \\
    c = \text{FCN}([c_0, c_1, c_2]), c \in \mathbb{R}^{F}, \\
    C = \text{Broadcast}(c), C \in \mathbb{R}^{F \times L}, \\
    \tilde{X}_m = X_m \odot C.
    \end{gathered}
\end{equation}
The frames in $X_m$ are passed through three Conv1d with different kernel sizes: $k_0$(small), $k_1$(middle), and $k_2$(large). Each is followed by an average pooling (AvgPool) and ReLU activation to deliver a weight vector $c_i$. Afterward, a two-layer down-up-sampled fully connected network (FCN) merges three weight vectors to create frequency-wise weights $c$. Then we broadcast the weights to operate element-wise multiplication with $X_m$ to get a weighted version $\tilde{X}_m$—accordingly, all the frames in the magnitude spectrogram share an identical frequency-dependent emphasis. 

\subsubsection{Modulator}

As mentioned earlier, we aim to explore a better conditioning method to remain in the same domain without encountering any domain conflicts. The mask estimator of SepFormer has shown its powerful ability to model sources from overcomplete features in the time domain. Hence, we build ConSep accordingly, viz., analyzing time-domain features, but additionally, the time signals are conditioned on the magnitude spectrogram. We rely on Feature-wise Linear Modulation (FiLM) \cite{film}. The FiLM allows adjusting the time signals to more appropriate representations based on the energy information of the given spectrogram. The residual connection is added after the FiLM layer to ensure the architecture performs well when the magnitude is relatively small. The modulating process can be formulated as:
\begin{equation}
    w = w_c+f_1(\Tilde{X}_m)\odot w_c+f_2(\Tilde{X}_m), \\
\end{equation}
where $w$ denotes the modulated feature, and each $f_i$ denotes an affine transformation.

\comment{
\begin{equation}
    \begin{gathered}
    f(\Tilde{X}_m) = \text{Tanh}(\mathcal{P}(\Tilde{X}_m)), \\
    g(\Tilde{X}_m) = \text{RReLU}(\mathcal{P}(\Tilde{X}_m)),
    \end{gathered}
\end{equation}
where $\mathcal{P}$ denotes a linear projection, $\text{Tanh}$ denotes hyperbolic tangent function, and $\text{RReLU}$ denotes randomized ReLU. The nonlinear functions are chosen for empirical reasons.
}

\subsection{Mask estimator}

%The overall structure of the mask estimator is displayed in the upper panel of Fig.~\ref{fig:sepformer}. 
The mask estimator inputs modulated feature $w$ and estimates a mask $m_k$. Firstly, the modulated feature $w$ is layer-wise normalized, chunked into overlapping segments with an overlap factor of 50\%, and then stacked. 

Afterward, the stacked feature feeds the SepFormer blocks, which exploit the dual-path mechanism \cite{dprnn}. The underlying process first captures the short-term dependencies by Intra-Transformer, then extracts the long-term dependencies by Inter-Transformer, and repeats $D$ times.
The unit Transformer structure used in Intra- and Inter-Transformer includes a multi-head attention (MHA) stage and a feed-forward (FFW) block with pre-LN setting \cite{pre-ln} and skip connections. Unlike DPTNet \cite{dptnet}, positional encoding is applied for injecting information on the order of sequence instead of a recurrent neural network (RNN). The total number of unit Transformers employed in Intra- and Inter-Transformer is $E$. 
A linear layer further processes the output of the SepFormer block to project the feature dimension for $K$ times deep. 
%The case with two active speakers is shown in the upper panel of Figure 3 with a bolder arrow.

\comment{
\begin{figure}[tb]
\begin{minipage}[b]{1.0\linewidth}
  \centerline{\epsfig{figure=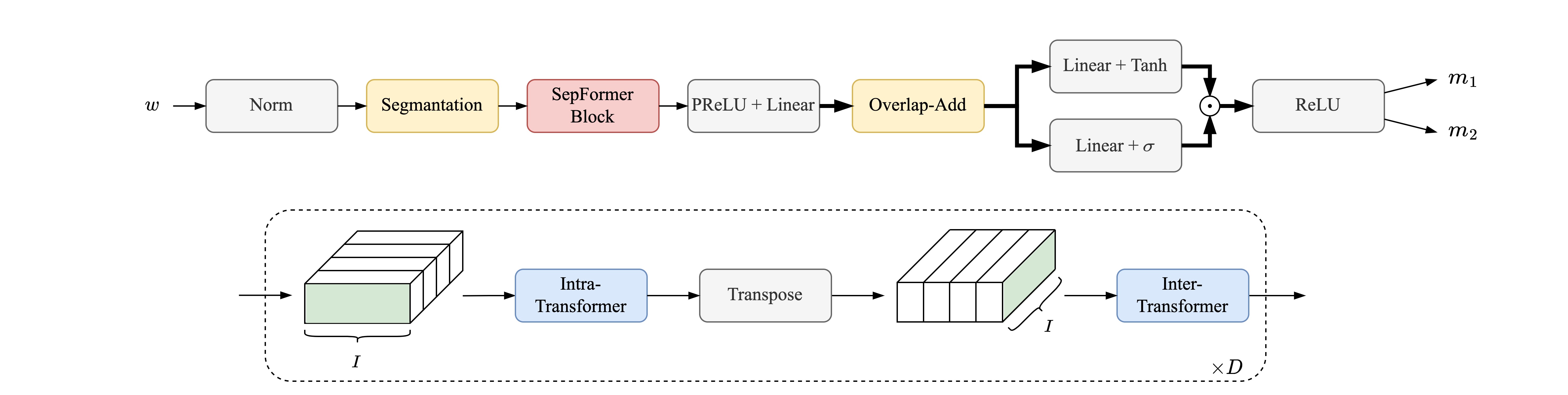, width=10cm}}
\end{minipage}
\caption{(Upper) The mask estimator of SepFormer without bottleneck projections. (Lower) The structure in SepFormer block, with the dual-path mechanism. The term $I$ indicates the input of the next layer.}
\label{fig:sepformer}
\end{figure} 
}

Finally, the projected output is passed through an overlap-add stage, two concurrent FFWs, and a ReLU activation to obtain the non-negative mask $m_k$. Note that we eliminate the bottleneck projections in the original SepFormer, as our prior experiment indicates its redundancy.

\subsection{Decoder}

The decoder is a transposed convolution layer with the same stride and kernel size as the learnable encoder. The input to the decoder for active speaker $k$ is the element-wise multiplication between the mask $m_k$ and the modulated feature $w$:

{\renewcommand\baselinestretch{0.8}\selectfont
\begin{equation}
\hat{s}_k = \text{Conv1d}^T (m_k \odot w).
\end{equation}
\par}

\section{Experimental setup}

\subsection{Datasets}

We validate the presented method on the popular WSJ0-2mix dataset \cite{dc} for the anechoic setting using the improvement of SI-SDR and SDR as the evaluation metrics. %In this dataset, the mixtures of two speakers are created by randomly mixing utterances in the WSJ0 corpus \cite{wsj}.
The training, validation, and test sets contain 30, 10, and 5 hours of speech data. The speech data are sampled at 8 kHz. %, and each mixture is clipped to the shorter component utterance’s length. Note that the training and test sets are from different speakers. 
Furthermore, we perform experiments in noisy settings. We rely on WHAM! \cite{wham} with urban noise and WHAMR! \cite{whamr}, which adds reverberation on top of WHAM!. These datasets are derived from WSJ0-2mix and have identical statistics.

\subsection{Training setup}

In the case of the learnable encoder, the encoder basis $N$ is set to 256. The input kernel size is 16 with a stride factor of 8. As for the STFT encoder, we use the Hamming window with a length of 256 (32 ms at 8 kHz), and the hop size is the same as that for the learnable encoder. For the MulCA, three kernel sizes $k_0$, $k_1$, and $k_2$ are 3, 5, and 10, respectively. 
Regarding the mask estimator, we follow the configuration proposed in \cite{sepformer1}, whereas $E$ is reduced to 4. 
For model training, we optimize the model using the Adam optimizer, a batch size of 1, and a learning rate of 1.5e-4. %, which is halved every 5 epochs if no improvement after 65 epochs in WSJ0-2mix. (80 and 100 for WHAM! and WHAMR!) We use automatic mixed-precision due to the computational limitations and the speed-up of training. 
Finally, the model is trained over 150 epochs with utterance-level Permutation Invariant Training (uPIT) \cite{pit} and SI-SDR losses. 

\section{Results and analyses}

\subsection{Comparison and Ablation study}

\begin{table}[tb]
\caption{Separation performance and speech quality metric on multiple conditions.}
\begin{center}
\label{tab:res}
\resizebox{8 cm}{!}{%
\begin{tabular}{c|ccc}
\hline
\textbf{Condition(s)}                       & \multicolumn{3}{c}{\textbf{Metrics}} \\ \hline
\textit{Anechoic} & \hspace*{1.5mm} \textit{SI-SDRi} \hspace*{1.5mm}   & \hspace*{1.5mm} \textit{SDRi}  \hspace*{1.5mm}   & \hspace*{1mm} \textit{NB-PESQ}  \\ \hline
SepFormer-s   & 16.53  & 17.02 & 3.12 \\
Bi-Sep32          & 16.28  & 16.49 & - \\
ConSep            & 16.72  & 17.19 & 3.39 \\ \hline
\textit{Noisy} & \textit{SI-SDRi} & \textit{SDRi} & \textit{NB-PESQ}  \\ \hline
SepFormer-s   & 13.23  & 14.13 & 2.50 \\
Bi-Sep32          & 13.62  & 14.54 & - \\
ConSep            & 13.82  & 14.82 & 2.76 \\ \hline
\textit{Noisy \& Reverberant} & \textit{SI-SDRi} & \textit{SDRi} & \textit{NB-PESQ}  \\ \hline
SepFormer-s   & 5.90   & 8.95 & 2.12 \\
Bi-Sep32          & 6.37   & 9.09 & - \\ 
ConSep            & 6.50   & 9.07 & 2.30 \\ \hline
\end{tabular}
}
\end{center}
\end{table}

\begin{table}[]
\caption{Ablation study. The third and fourth columns use different conditioning methods without MulCA as well.}
\label{tab:ablation}
\resizebox{\columnwidth}{!}{%
\begin{tabular}{c|ccccc}
 &
  ConSep &
  w/o MulCA &
  \begin{tabular}[c]{@{}c@{}}w/o FiLM\\ (concat+linear)\end{tabular} &
  \begin{tabular}[c]{@{}c@{}}w/o FiLM\\ (add)\end{tabular} &
  \begin{tabular}[c]{@{}c@{}}w/o \\ conditioning\end{tabular} \\ \hline
SI-SDRi &
  13.82 &
  13.72 &
  13.59 &
  13.34 &
  13.23 \\
SDRi &
  14.82 &
  14.68 &
  13.96 &
  14.19 &
  14.13
\end{tabular}%
}
\end{table}

We compare the separation accuracy of ConSep with SepFormer-s and Bi-Sep as baselines. Tab.~\ref{tab:res} presents the experimental results in terms of both SI-SDRi and SDRi. It demonstrates the advantages of our proposed model. For all kinds of environments, ConSep outperforms all other methods except the SDRi, which can be deceived by the loudness \cite{sisdr}, in noisy and reverberant settings. Moreover, the fact that a better conditioning strategy not only gains stability but succeeds to the beneficial sides is proven, as ConSep surpasses Bi-Sep in non-anechoic settings and SepFormer in the anechoic setting.

Furthermore, we show the evaluation in speech quality metric. We can observe that employing the STFT features may improve narrow-band PESQ (NB-PESQ), and the same phenomenon can be observed in \cite{compensate, explore, sesep}. 
This adds another clue of using a larger frame time-frequency representation still presenting as a desirable feature, even if learned-domain models have more freedom to adapt to the SI-SDR training loss.

To validate the effectiveness of ConSep, we also conduct an ablation study in noisy environments, as shown in Tab.~\ref{tab:ablation}. We validate the performance without MulCA and magnitude conditioning. Additionally, we experiment with various conditioning methods, including simply adding or concatenation followed by a down-sizing linear layer. The result shows that with attention to magnitudes, the separation performance improves, as "concat+linear" performs closely to Bi-Sep. However, employing FiLM brings the most apparent improvement among other methods. (see the entry "w/o MulCA") 

\subsection{Visualization}

\begin{figure}[tb]
\begin{minipage}[b]{1.0\linewidth}
  \centering
  \centerline{\epsfig{figure=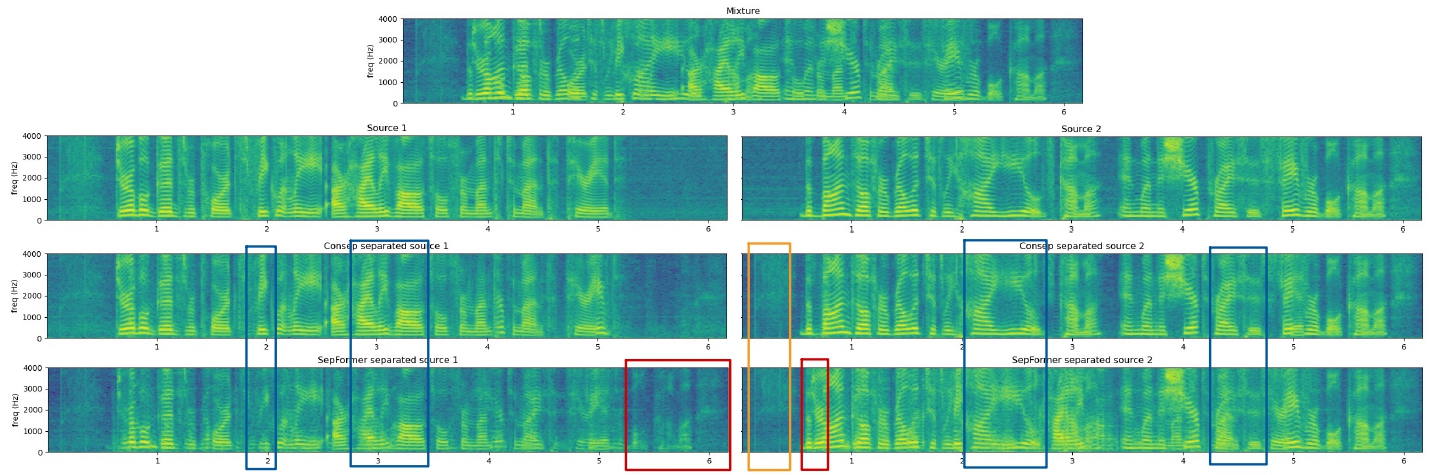, width=8.9cm}}
   \centerline{\small{(a) \textit{Anechoic}}}\medskip
\end{minipage}
\begin{minipage}[b]{1.0\linewidth}
  \centering
  \centerline{\epsfig{figure=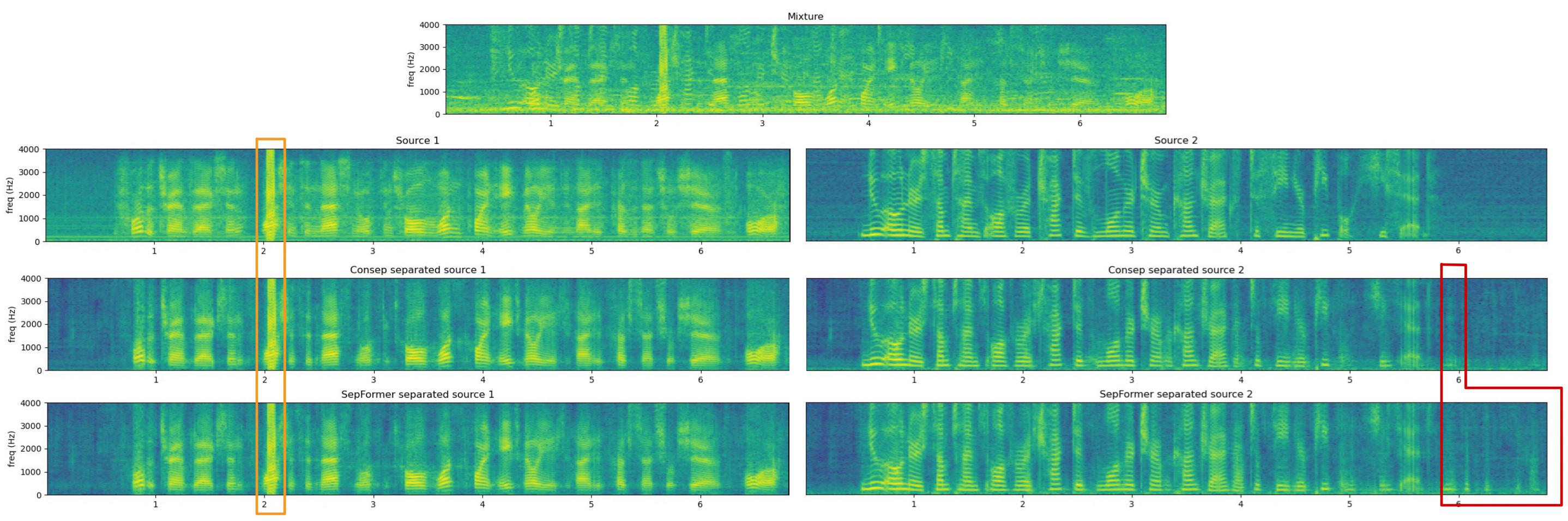, width=8.9cm}}
  \centerline{\small{(b) \textit{Noisy}}}\medskip
\end{minipage}
\begin{minipage}[b]{1.0\linewidth}
  \centering
  \centerline{\epsfig{figure=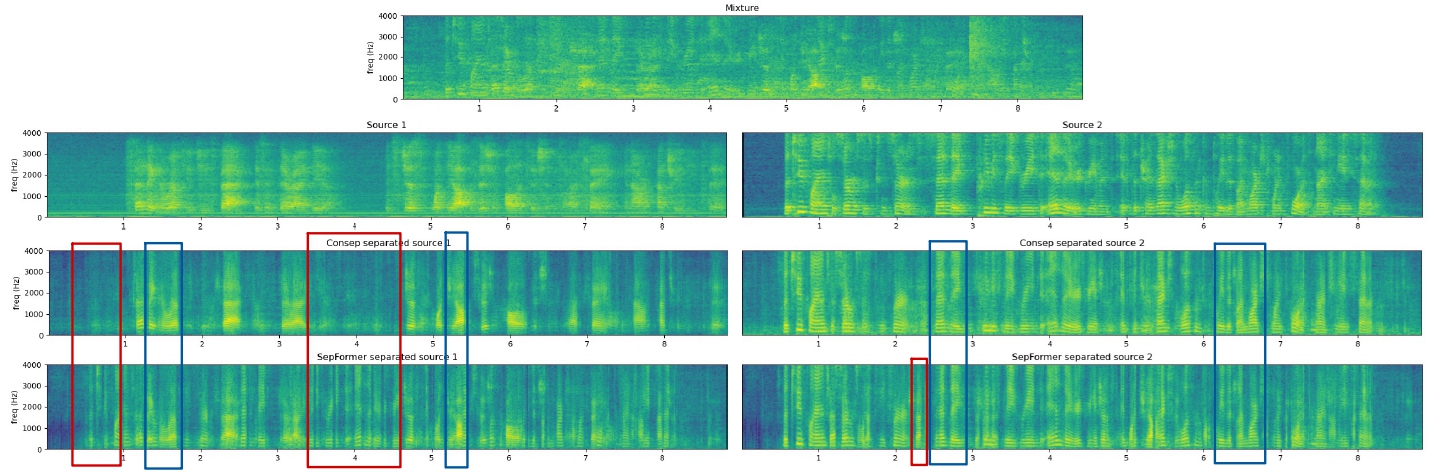, width=8.9cm}}
  \centerline{\small{(c) \textit{Noisy \& Reverberant}}}\medskip
\end{minipage}
\caption{Case studies. Generally, the rows indicate the spectrogram of mixture, sources, ConSep output, and SepFormer output from top to bottom. The two columns indicate the first and second sources from left to right. Also, red and blue boxes denote false alarm issues and spectral/harmony clarity. For (a) and (b), the non-speech signals cropped in the orange box are the sounds of inhaling and microphone pop, respectively.}
\label{fig:case}
\end{figure} 

\begin{figure}[t]
\begin{minipage}[b]{1.0\linewidth}
  \centering
  \centerline{\epsfig{figure=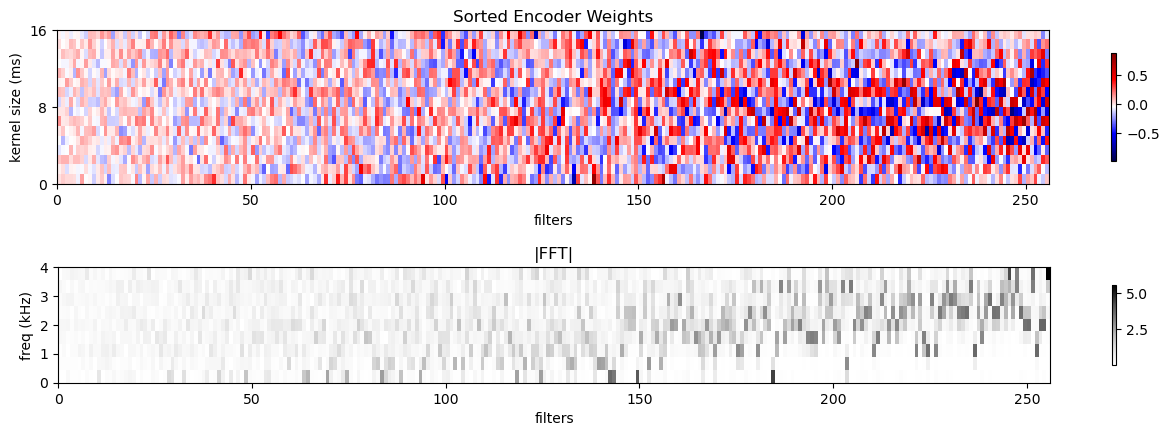, width=8.5cm}}
  \centerline{\small{(a) ConSep}}\medskip
\end{minipage}
\begin{minipage}[b]{1.0\linewidth}
  \centering
  \centerline{\epsfig{figure=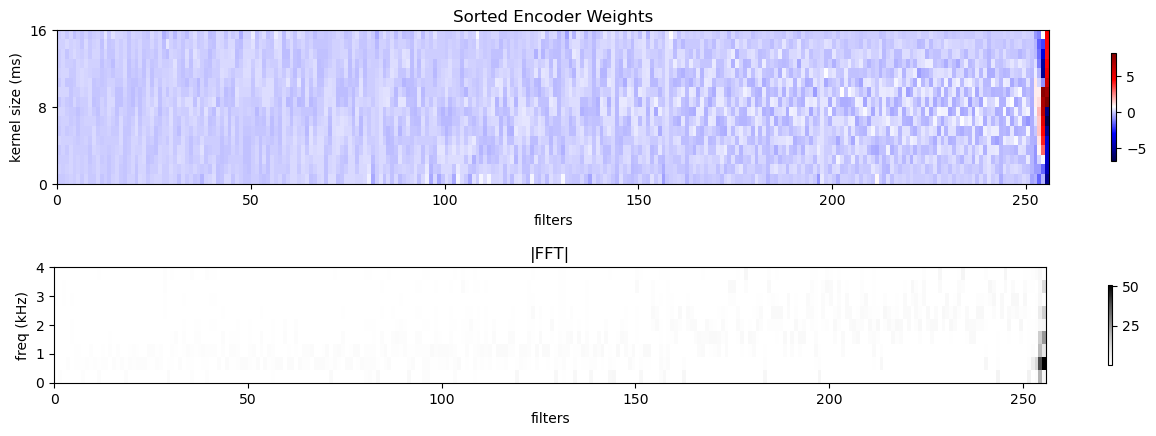, width=8.5cm}}
  \centerline{\small{(b) SepFormer-s}}   \medskip
\end{minipage}
\caption{For each sub-figure, the upper and lower panel depict the encoder bases sorted by Euclidean similarity and their frequency response, respectively.}
\label{fig:two-enc}
\end{figure} 

We visualize the mixture, clean sources, and separated outputs from ConSep and SepFormer, as shown in Fig.~\ref{fig:case}. 
\comment{For all the sub-figures in Fig.~\ref{fig:case}, the red and blue boxes illustrate issues of false alarm and spectral/harmony clarity, respectively. Additionally, we highlight some non-speech signals inherited from sources in orange boxes.}
The mixture in Fig.~\ref{fig:case}(a) consists of two women speaking with a similar pitch range. We can see that the frequency contour is more prominent in ConSep-separated sources and that SepFormer tends to produce false alarms. This implies that merely analyzing time signals from an overcomplete set of encoder bases renders the model confusing when facing speakers with similar pitch identities. As for Fig.~\ref{fig:case}(b), the mixture consists of a woman, a man speaking, and cafeteria noises. We can see that both models denoise well, but likewise, some speech-related false alarms occur. The same observation can also be pointed out in Fig.~\ref{fig:case}(c), whose mixture consists of the reverberant version of two men speaking with poll hall noises. Regardless, the superiority of a more apparent contour in ConSep remains, probably due to the enhancement by attending the spectrogram that better presents the harmonics. This concludes that the essential components of speech signals are better captured in ConSep. 

Furthermore, we plot the sorted bases of the learnable encoder trained in anechoic condition and their frequency response, as shown in Fig.~\ref{fig:two-enc}. Resembling \cite{convtasnet}, most filters are tuned to lower frequencies. This suggests an essential role for low-frequency speech features such as pitch to achieve better performance. However, we notice a few weird filters on the right side of Fig.~\ref{fig:two-enc}(b), which can translate to the high-valued low-pass filter at the end of the frequency response. This may be the reason why SepFormer could not perform well when attention to high-frequency information is required, such as the circumstances when facing similar-pitch speakers.

\section{Conclusions}

In this study, we propose a noise- and reverberation-robust speech separation framework, ConSep, by means of conditioning the time signals by magnitude spectrogram. The goal of generalizability has been fulfilled as this framework upgrades an existing model to fit various environments. Furthermore, we analyze and visualize the results to get a better picture of the advantages of ConSep, which as well demonstrates phenomena coherent with the actualities found through preliminary studies.

\vspace{12pt}

\end{document}